\documentclass[12pt]{article}
\usepackage{graphicx}
\begin{document}
version 9.11.2011
\begin{center}
{\Large On the uniqueness of the Standard Model of particle physics}  
\vspace{0.3cm}

H.P. Morsch\footnote{postal address: Institut f\"ur Kernphysik,
Forschungszentrum J\"ulich, D-52425 J\"ulich, Germany\\  E-mail:
h.p.morsch@gmx.de }\\
National Centre for Nuclear Research, Pl-00681 Warsaw, Poland 
\end{center}

\begin{abstract}
The running strong coupling $\alpha(Q)_s$ and the gluon propagator
from QCD have been compared to similar quantities in the quanton model,
a generalisation of QED with massless fermions (quantons) and scalar
coupling of boson fields. In the latter model a series of bound
states (which can be related to different flavours) have been
obtained. Assuming a weighting of their momentum distributions with the
average momentum $\tilde Q_i$ of each state, a running of the coupling
$\alpha(Q)$ is obtained, which is in quantitative agreement with
$\alpha_s(Q)$ from QCD. Also with a similar weighting the
gluon propagator from lattice QCD simulations is well described. This
indicates clearly that QCD and thus the Standard Model is not a unique
description of fundamental interactions. 

Different from the Standard Model, the quanton model is simple
and yields bound states with correct masses. This may
indicate that this model yields a more realistic description of
fundamental forces.

PACS/ keywords: 11.15.-q, 12.40.-y, 14.40.-n/ Comparison of
the running coupling $\alpha(Q)$ and the gluon propagator of QCD with
the quanton model, an extension of QED with scalar coupling of gauge
bosons. Equivalent description in both models indicating that the
Standard Model is not unique. 
\end{abstract}

From the ultimate theory of fundamental forces one expects a
structure, which is simple, complete and unique. Whether nature
reduces the fundamental forces to a very simple form is not known and
has to be found out, but completeness can be tested. Apart
from the fact that the theory has to describe the wealth of existing
data, it is important that it couples to the absolute vacuum
with energy $E_{vac}=0$. The uniqueness can be tested only by
comparing the theory with other models.

The Standard Model of particle physics~\cite{PDG} (SM) is composed of
different gauge theories, quantum electrodynamics (QED), weak
interaction theory, and quantum chromodynamics (QCD), and has
been applied successfully to the description of hadronic and leptonic
processes over the last decades. 
It is certainly not simple and also not complete, since the elementary
fermions (quarks and leptons) are massive. Another shortcoming of the
SM is that in the relativistic theory stationary states of 
elementary fermions, the most important features of particle
physics, cannot be described. This has been taken less seriously,
since bound states can be described in non-relativistic 
approximations. However, this is not possible for the strong
interaction, which has a large coupling strength. 

To understand the mass problem of elementary fermions in the SM, 
the Higgs-mechanism has been adopted~\cite{Higgs}, in which the 
g.s.~of elementary particles is lowered by symmetry breaking. However,
this mechanism implies the existence of heavy Higgs-bosons, which have never
been found in spite of extensive searches over 20 years. Another
severe problem of the Higgs-mechanism is that it requires an
enormously high energy density of the vacuum in striking disagreement
with astrophysical observations. 

Recently, a new formalism (quanton model) has been
developed~\cite{Moneu1} for the description of fundamental forces,
based on a different understanding of mass, given solely by 
binding effects of elementary massless\footnote{with insignificant
  mass with respect to the mass scale of particle physics} fermions
(quantons). This model can be used to test the uniqueness of the SM.  
In the present paper two basic quantities of QCD - which have been
investigated extensively - have been compared with corresponding
properties of the quanton model: the running strong coupling
$\alpha_s(Q)$ and the 'gluon propagator', the latter taken from
lattice gauge simulations~\cite{gprop,gLandau,gLaplace,gCoulomb}. 

Starting from QED with scalar coupling of two boson fields
$A_{\mu}A^{\mu}$, the Lagrangian may be written in the form 
${\cal L}=\frac{1}{\tilde m^{2}} \bar \Psi\ i\gamma_{\mu}D^{\mu}(
D_{\nu}D^{\nu})\Psi\ -\ \frac{1}{4} F_{\mu\nu}F^{\mu\nu}$,   
where $\tilde m$ is a mass parameter, $\Psi$ a massless fermion
(quanton, q) field, $D_{\mu}=\partial_{\mu}-i{g} A_{\mu}$ the
covariant derivative, and $F^{\mu\nu}=\partial^{\mu}A^{\nu}
-\partial^{\nu}A^{\mu}$ the field strength tensor. In this Lagrangian
higher derivatives of the fermion field appear~\cite{highd}, which 
have to be removed. This is possible~\cite{Moneu1} by a symmetry between
fermions and one boson field  $\partial^{\nu}-i{g} A^{\nu}= const$, leading
to a constrained Lagrangian of the form    
\begin{equation}
\label{eq:Lagra}
{\cal L}=\frac{1}{\tilde m} \bar \Psi\ i\gamma_{\mu}D^{\mu}
D_{\nu}\Psi\ -\ \frac{1}{4} F_{\mu\nu}F^{\mu\nu}~, 
\end{equation}
The first term of this Lagrangian gives rise to the following two terms  
\begin{equation}
\label{eq:L2}
{\cal L}_{2g} =\frac{-ig^{2}}{\tilde m^{2}} \ \bar \Psi \gamma_{\mu}
A^{\mu} (\partial_\nu A^\nu) \Psi \ 
\end{equation}
and 
\begin{equation}
\label{eq:L3}
{\cal L}_{3g} =\frac{ -g^{3}\ }{\tilde m^{2}} \ \bar \Psi \gamma_{\mu}
A^\mu (A_\nu A^\nu)\Psi \ .  
\end{equation}
These two parts of the Lagrangian  do not have 
derivatives of the fermion field. Further, the derivative of $A^\nu$
appears in ${\cal L}_{2g}$, but this field appears in ${\cal L}_{3g}$
only in combination with another boson
field. Therefore, the problems of higher derivative Lagrangians are
avoided. Generally, a Lagrangian with a mass denominator $\tilde m^n$ (with $n\geq
1$) leads to a non-renormalizable theory, however, in the present case
the Lagrangian yields finite matrix elements and consequently 
results in full space.  

From these Lagrangians ground state matrix elements $<g.s.|
~K(\tilde p'-\tilde p)~|g.s.>$ have been derived, where $K(\Delta \tilde
p)$ are the multi-boson field operators of the above
Lagrangians~(\ref{eq:L2}) and (\ref{eq:L3}). These yield contributions
only, if the two boson fields overlap in space and time. By equal time
requirement of the overlapping boson fields a reduction to three
dimensions is possible, which gives rise to two potentials, which are
given in r-space by
\begin{equation}
V_{2g}(r)= \frac{\alpha^2\hbar^2 \tilde E^2}{2\tilde m^3}\ \Big
(\frac{d^2 w(r)}{dr^2} + 
  \frac{2}{r}\frac{d w(r)}{dr}\Big )\frac{1}{\ w(r)}\ ,  
\label{eq:vb}
\end{equation}
and 
\begin{equation} 
\label{eq:vqq}
V_{3g}(r)= \frac{\hbar}{\tilde m} \int dr'\rho(r')\ V_{g}(r-r')~,  
\end{equation}
where $w(r)$ and $\rho(r)=w^2(r)$ are wave function and (quasi)
density\footnote{$\rho(r)$ has dimension $fm^{-2}$ due to the
  dimension $GeV^2$ of the two-boson field.} of
a two-boson field, $\tilde E$ the mean energy of scalar
excitation in the potential~(\ref{eq:vb}), and $V_{g}(r)$ a
boson-exchange potential given by $V_{g}(r)=-\alpha^3\hbar \frac{f(r)}{r}$. 

The potential $V_{2g}(r)$ corresponds to the 'confinement' potential
required in hadron potential models~\cite{qq}, whereas  $V_{3g}(r)$ is
related to the usual boson-exchange potential  
derived from basic gauge theories, but due to its more complex
structure it is finite for $r\to 0$ and $\infty$ and scales with the
coupling strength $\alpha^3$.

For a $q\bar q$ system in a scalar state ($J^{\pi}=0^+$), angular momentum
L=1 is needed. Therefore, for this case a p-wave density is required in
eq.~(\ref{eq:vqq}), which is related to $\rho(r)$ by 
\begin{equation}
\label{eq:spur}
\rho^{ p}(\vec r)=\rho^{ p}(r)\ Y_{1,m}(\theta,\Phi) =
(1+\beta R\ d/dr) \rho(r)\ Y_{1,m}(\theta,\Phi)\ .  
\end{equation}
$\beta R$ is determined from the condition $<r_{\rho^p}>\ =\int
d\tau\ r \rho^p(r)=0$ (elimination of spurious motion). 
This yields a boson-exchange potential given by
\begin{equation}
\label{eq:vqq0}
V^{s}_{3g}(r)= \frac{\hbar}{\tilde m} \int d\vec r\ '\rho^{ p}(\vec r\ ')\
Y_{1,m}(\theta',\Phi')\ V_{g}(\vec r-\vec r')
= 4\pi \frac{\hbar}{\tilde m} \int dr'\rho^{ p}(r')\ V_{g}(r-r')~. 
\end{equation}

By requiring that the boson-exchange force can act only within the
two-boson density $\rho(r)$, which gives rise to the constraint
\begin{equation}
\label{eq:con1}
V^{s}_{3g}(r)=c_{pot} \ \rho(r)\ ,
\end{equation}
the density as well as the parameters of
the interaction cut-off function
$f(r)=(e^{(ar)^{\sigma}}-1)/(e^{(ar)^{\sigma}}+1)~e^{-cr}$ can be  
deduced self-consistently, see the details in ref.~\cite{Moneu1}.  
This yields  
\begin{equation}
\label{eq:wf}
\rho(r)=\rho_o\ [exp\{-(r/b)^{\kappa}\} ]^2\ \ with\ \ \kappa \sim
1.3-1.5\ . 
\end{equation} 

Inserting this form of $\rho(r)$ in $V_{2g}(q)$~(\ref{eq:vb}) leads to
the explicite form
\begin{equation}
V_{2g}(r)= \frac{\alpha^2\hbar^2 \tilde E^2}{\tilde m^3}\
\Big[\frac{\kappa}{b^2} (\frac{r}{b}) 
^{\kappa-2}\ [\kappa (\frac{r}{b})^{\kappa} - (\kappa +1)]\Big]\ .
\label{eq:vbind}
\end{equation}

The mass of the system is defined by  
\begin{equation}
\label{eq:mass}
M_{n}=-E_{3g}+E_{2g}^{~n} \ , 
\end{equation}
where $E_{3g}$ and $E_{2g}^{~n}$ are the binding energies in
$V^{s}_{3g}(r)$ or $V^{v}_{3g}(r)$ and $V_{2g}(r)$, respectively,
calculated by using a mass parameter $\tilde m={1}/{2}~\tilde M$,
where $\tilde M$ is the average mass of the system, weighted over vector
and scalar states. However, since this weighting is not known, 
$\tilde m={1}/{2}~M$ is used, where $M$ is the ground state mass of the
system. This allows to use the additional constraint $M=M_1$. In this
way, the mass contributions due to excited states is included in $\tilde E$,
which is used as fit parameter. The coupling constant $\alpha$ is obtained 
by matching the mass to the lowest binding energy in eq.~(\ref{eq:mass}). 
The binding energies in $V_{3g}(r)$ are negative. Using the energy-momentum relation
in the form $E_{vac}=0=\sqrt{<Q^2_\rho>}+\tilde E_{3g}$, where $<Q^2_\rho>$ is the
mean square momentum of $\rho(r)$ and $\tilde E_{3g}$ a weighted
average of $E_{3g}$ between vector and scalar states, this yields another constraint  
\begin{equation}
\label{eq:bindm}
\tilde E_{3g}=-{\sqrt{<Q^2_\rho>}}~. 
\end{equation}
Different from the binding in $V_{3g}(r)$, which does not correspond
to real mass generation, the binding energy $E_{2g}$ is
positive and allows creation of stable particles out of the absolute
vacuum of fluctuating boson 
fields, if two rapidly fluctuating boson fields overlap and trigger
a quantum fluctuation with energy $E_{2g}$. 

Using the constraints~(\ref{eq:con1}) and (\ref{eq:bindm}) and the
energy-mass relation~(\ref{eq:mass}), all parameters of the model are
determined (with some ambiguities) for a given slope parameter b. This
last parameter may be determined by the structure of the vacuum. For
mesonic systems this is discussed in ref.~\cite{Moneu2} and leads to a
description, in which all parameters of the model are fixed.  
This is based on a vacuum potential sum rule, assuming a global
boson-exchange interaction in the vacuum $V_{vac}(r)\sim
1/r^2$. Further, the different potentials $V^i_{3g}(r)$ (where $i$ are
the discrete solutions) sum up to $V_{vac}(r)$ 
\begin{equation}
\label{eq:sum}
\sum_i V^i_{3g}(r)=V_{vac}(r)= \tilde f_{as}(r) (-\tilde \alpha_e^3
\hbar\ r_o/{r^2})\ e^{-\tilde cr} \ , 
\end{equation}
where $\tilde f_{as}(r)$ and $e^{-\tilde cr}$ are cut-off functions
similar to those for the boson-exchange interaction discussed above.

A sum rule analysis similar to that discussed in ref.~\cite{Moneu2} has
been performed, but for the present study the higher energy region up
to several hundred GeV has been included. A comparison of the
resulting boson-exchange potentials with the sum 
rule~(\ref{eq:sum}) is given in fig.~1 with the deduced masses and
 parameters in table~1. In addition to the states
discussed in ref.~\cite{Moneu2} another solution is obtained in the
hundred GeV region, with a vector state at about 91 GeV and a scalar
state at the mass of the observed $t\bar t$ system; therefore, these two states
can be identified with a 'top' flavour system. Interestingly, the
vector state is at the mass of the $Z^o$ boson, consequently, this
particle has to be identified in our model as part of the 'top' system.

\begin{table}
\caption{Deduced masses (in GeV) of scalar and vector $q^+q^-$
  states in comparison with the lowest $0^{++}$ and $1^{--}$
  mesons~\cite{PDG}.}   
\begin{center}
\begin{tabular}{lcc|c}
Solution&(meson)& $M$ & $M^{exp}$ \\ 
\hline
1~~~scalar& $\sigma$ & 0.55 & 0.60$\pm$0.2  \\ 
\hline
2~~~scalar&$ f_o  $ & 1.70 & 1.70$\pm$0.2 \\ 
~~~~vector&$\omega$ & 0.78 & 0.78 \\ 
\hline
3~~~scalar&$ f_o  $ & 3.28  \\ 
~~~~vector&$\Phi$   & 1.02 & 1.68$\pm$0.02\\ 
\hline
4~~~scalar& not seen & 12.7  \\ 
~~~~vector&$J/\Psi$  & 3.10 & 3.097   \\ 
\hline
5~~~scalar& not seen & 40.4  \\ 
~~~~vector&$\Upsilon$ & 9.46 & 9.46 \\
\hline
6~~~scalar& top & $\sim$370   & $\sim$370 \\ 
~~~~vector& (Z$^o$) & $\sim$91 & 91.2 \\
\end{tabular}
~ \\
\begin{tabular}{c|cc|ccc|ccc}
Sol.&$\kappa$& $b$ &$\alpha_e$& $c$ &$a$ &
$<r^2_{\rho}>$ & $<Q^2_{\rho}>^{1/2}$ &  $\tilde E$ \\   
\hline
1~& 1.50  & 0.831  & 0.257 &  2.24 & 5.8 & 0.761 & 0.59  & 1.0   \\ 
2~& 1.46  & 0.264  & 0.260 &  7.20 & 18  & 0.080 & 1.57  & 0.6  \\ 
3~& 1.44  & 0.149  & 0.281 &  13.2 & 30  & 0.026 & 2.98  & 0.5  \\ 
4~& 1.40  & 0.054  & 0.327 &  33   & 82  & 0.36 10$^{-2}$ & 7.73  & 0.8  \\ 
5~& 1.36  & 0.020  & 0.390 &  95   & 220 & 0.51 10$^{-3}$ & 20.7  & 1.5  \\ 
6~& 1.30  & 0.0047 & 0.714 &  390  & 900 & 0.32 10$^{-4}$ & 90.8  & 4.4  \\ 
\end{tabular}
\end{center}
\end{table}

To make a comparison with the running coupling strength
$\alpha^{QCD}(Q)$ one has to go to the Fourier transform of the vacuum
potential $V_{vac}(r)$~(\ref{eq:sum}), which can be written in the
form $V_{vac}(Q) =\alpha^3(Q)/Q =\sum_i V^i_{3g}(Q)$. This leads to
$\alpha(Q)=[\sum_i V^i_{3g}(Q)]^{1/3}\ Q^{1/3}$. Individual coupling
functions $\alpha_i(Q)$ can also be defined by $\alpha_i(Q)=
[V^i_{3g}(Q)]^{1/3}\ Q^{1/3}$. For the comparison with QCD the
following form is used
\begin{equation}
\label{eq:aqcd}
\alpha^{QCD}(Q)=[\sum_i w_iV^i_{3g}(Q)]^{1/3}\ Q^{1/3}   
\end{equation}
with weighting factors $w_i$, which are adjusted to the QCD data in
fig.~2. A quantitative agreement with $\alpha^{QCD}(Q)$ is obtained in
the momentum range covered by the solutions in table~1 assuming
$w_i=\frac{1}{3}<Q^2_i>^{-1/2}$. This is shown in fig.~4, which
dispays eq.~(\ref{eq:aqcd}) by solid line as well as 
the individual components by dashed and dot-dashed lines.

From this comparison one can draw the following conclusions: first, the
correspondence with QCD supports the concept of a vacuum potential sum 
rule in the present model with a sequence of solutions consistent with different
flavour states. It is evident that $\alpha(Q)$ is nothing else but a
different form of the vacuum potential sum rule with a strong
weighting of the high momentum region. One can see that the 'top'
system at about 100 GeV is needed. Further, the abrupt fall-off beyond
$Q$=100 GeV in our calculations indicates that the flavour states are
not limited to the states presently known but continue to larger
masses, with the next flavour state in the one TeV mass region.  
Second, the factor 1/3 in $w_i$ is probably due to colour, whereas the
factor $\tilde Q=<Q^2_i>^{-1/2}$, which leads to the strong fall-off
of $\alpha^{QCD}(Q)$, appears to be due to the non-Abelian character of QCD. 
If one relates $\alpha(Q)$ to the mass, the quanton model is correct,
whereas $\alpha^{QCD}(Q)$ suffers from increasing mass deficits for
larger Q-values, apparently related to the strong increase in the
needed quark masses.

Another problem of QCD is that only perturbative solutions exist,
restricting QCD analyses to reactions with large momentum transfers,
as deep inelastic scattering. For this type of reaction, the extracted
strength has to be weighted with the probability of the momentum
transfer $Q$, which is 
$1/Q$. Therefore, the quanton model yields the same result as QCD, if
for deep inelastic reactions the strength is weighted with $1/Q$.  

For a second comparison with QCD at much smaller momenta, one has to
go to lattice gauge theory, which is not limited to perturbative solutions.
In this approach hadron masses have been calculated quite
successfully, but also a good description of the confinement
potential~\cite{Bali} has been obtained. In particular, the 'gluon
propagator' has been investigated quite intensely, see
e.g.~refs.~\cite{gprop,gLandau,gLaplace,gCoulomb}, since its structure 
lacks a simple direct understanding. This quantity is not gauge
independent, a summary of lattice calculations in different
gauges~\cite{gLandau,gLaplace,gCoulomb} is 
shown in fig.~5. In our framework we describe the corresponding 'boson
propagator' $P_g$ as a sum of boson-exchange potentials over i
flavour states~\cite{Moneu2} $P_g=\sum_i  V_{3g}^i(Q)$. Similar to the
analysis of $\alpha^{QCD}(Q)$ we replace $P_g$ by 
\begin{equation}
\label{eq:prop}
P^{QCD}_g=\sum_i w_i' V_{3g}^i(Q)
\end{equation}
with weighting factors $w_i'$ fitted to the lattice data. Good fits are shown in
fig.~5 with average weightings $w_i'\sim33<Q_i^2>^{-1/2}$ and individual
contributions given by the dot-dashed lines. So, we observe a similar
$<Q_i^2>^{-1/2}$ dependence as found for $\alpha^{QCD}(Q)$. 

Concerning the weak interaction part of the SM, this interaction may
be replaced in the quanton model by the spin-spin interaction between
quantons, assuming a structure of leptons as bound states of three
quantons, see ref.~\cite{Moneu3}. It is interesting to note that the
spin-spin force for neutral elementary particles is not considered in
the SM. This interaction has properties very similar to the weak
interaction, a coupling strength many orders of magnitude smaller
than the electromagnetic force and consequently an extremely short
range~\cite{Moneu3}. However, the crucial difference to the weak
interaction theory is that the spin-spin force does not require
massive gauge bosons and (due to the Higgs-mechanism) a large energy
density of the vacuum.  
\vspace{0.3cm}
 
In conclusion, calculations of the momentum dependence of the coupling
$\alpha(Q)$ and the boson propagator within the quanton model have
been compared with the corresponding quantities from QCD. A 
good agreement of these quantities is obtained in both theories, if
the results of the quanton model are scaled with the momentum. This
indicates clearly that QCD and thus the SM is not the unique theory of
the strong interaction. 

On the other hand, the quanton model appears to have all necessary properties of a
fundamental theory, it has a very simple structure with only two quantons,
charged and uncharged, coupled by the same gauge boson to the
charge and spin, respectively. Further, it may to be complete, but
further test are needed before this model can be regarded as a
realistic description of fundamental forces. 
\vspace{0.5cm}

We thank B.~Loiseau for his continuous support during the development of the model. 
\vspace{0.3cm}


\newpage
\begin{figure}
\centering
\includegraphics [height=18cm,angle=0] {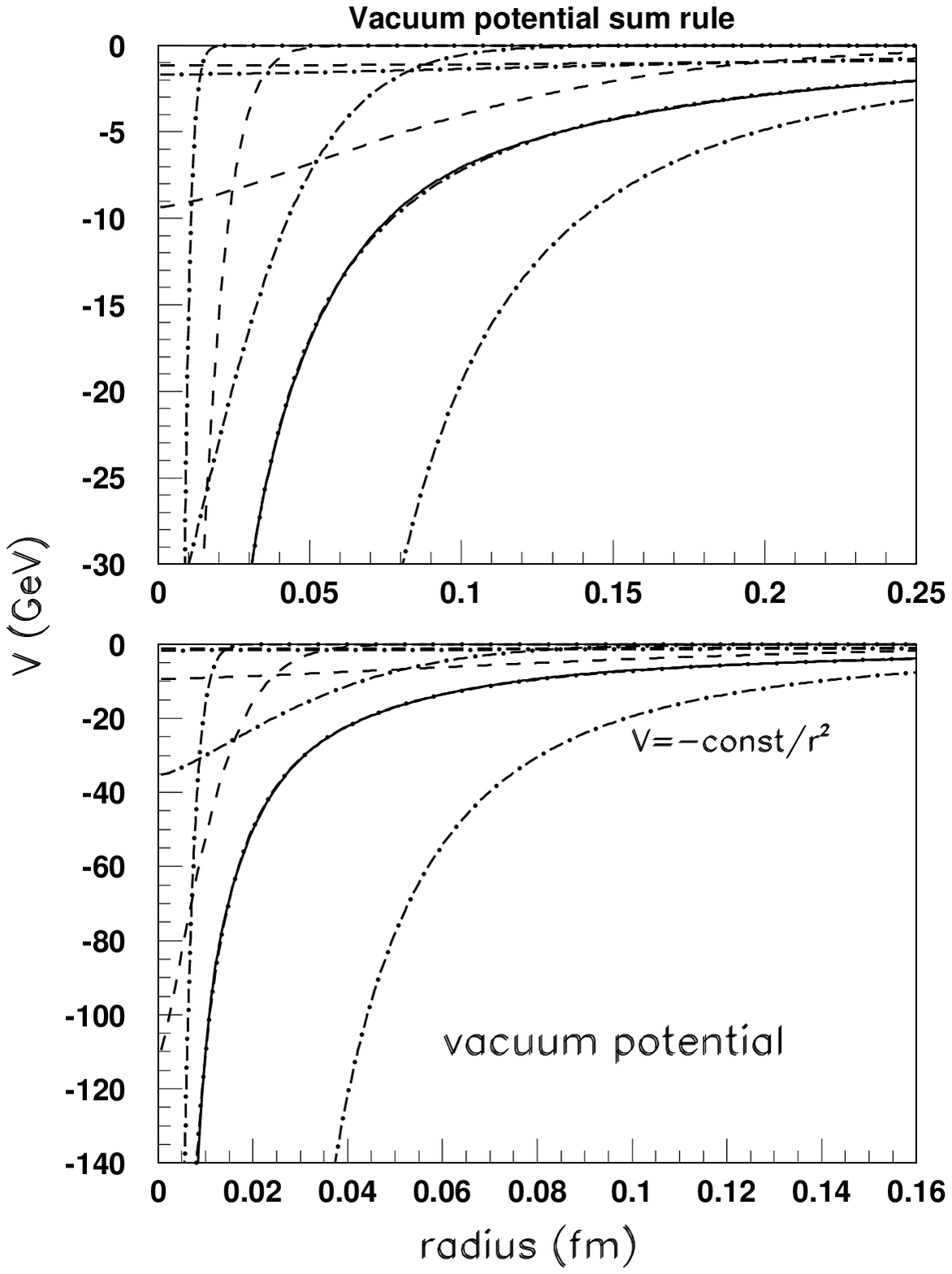}
\caption{Boson-exchange potentials for different solutions (given by
  dot-dashed and dashed lines) and sum given by solid line. This is
  compared to the vacuum sum rule~(\ref{eq:sum}) given by the
  dot-dashed line overlapping the solid line. A pure potential
  $V=-const/r^2$ is shown also by the lower dot-dashed line. The lower
  part shows the same lines but with a vertical scale enlarged to
  -140 GeV.} 
\label{fig:sumvqqall}
\end{figure} 

\begin{figure} [ht]
\centering
\includegraphics [height=16cm,angle=0] {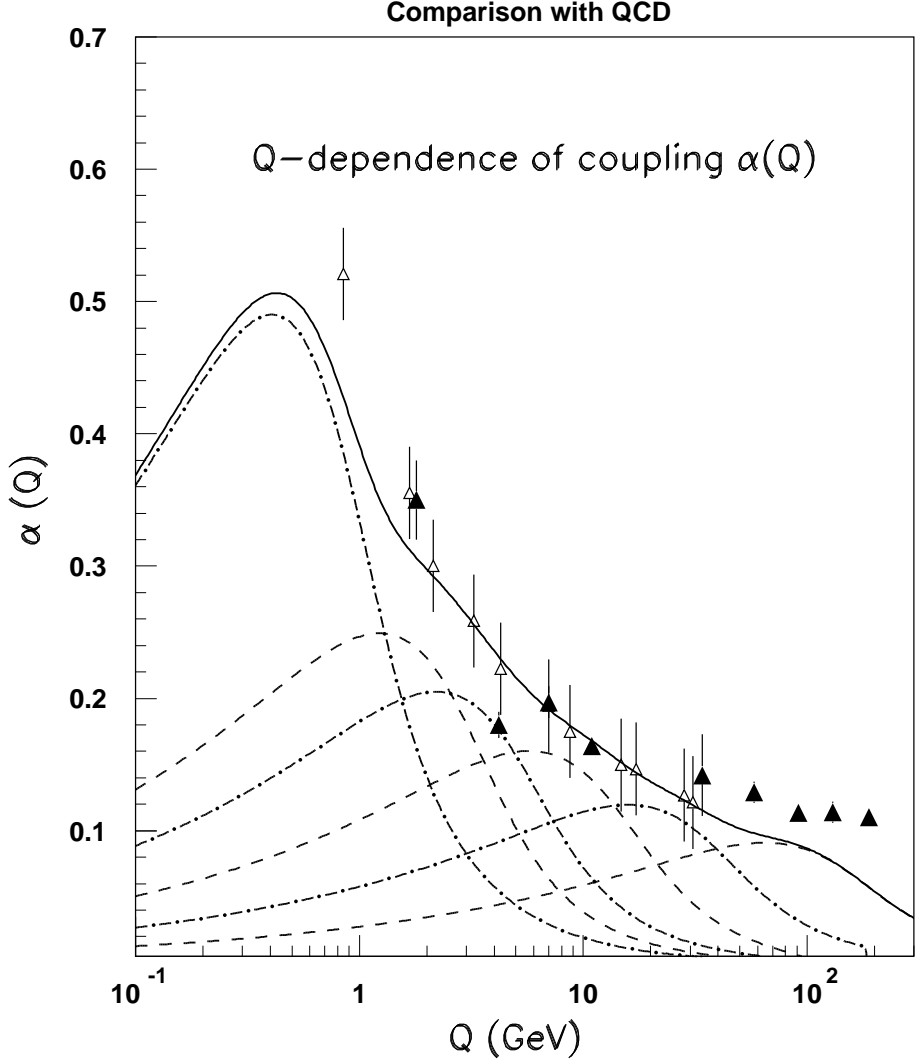}
\caption{Momentum dependence of the coupling strength
  $\alpha(Q)$ from QCD analyses (solid triangles) and lattice QCD
  simulations (open triangles) in comparison with 
  our results. Applying for all solutions i a normalisation
  $\frac{1}{3} <Q^2_i>^{-1/2}$, the dot-dashed and dashed lines
  correspond to the individual solutions~\cite{Moneu2} with
  an additional solution in the 100 GeV region. The sum is given by
  the solid line, which is in good agreement with QCD. }   
\label{fig:g1vac}
\end{figure}

\begin{figure} [ht]
\centering
\includegraphics [height=18cm,angle=0] {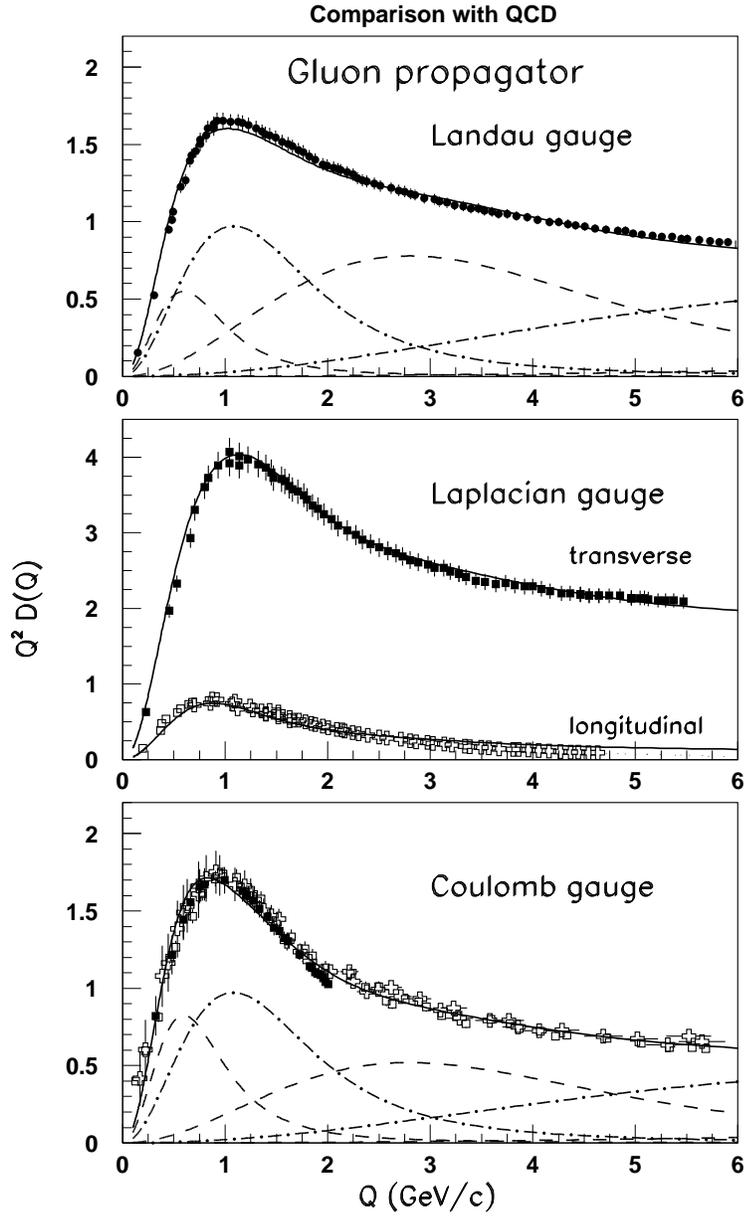}
\caption{Data on the gluon propagator from lattice gauge calculations
  in different gauges~\cite{gLandau,gLaplace,gCoulomb} in comparison
  with calculations within our 
  model using a sum of the different flavour contributions given in
  ref.~\cite{Moneu2}, which yields a good description of the lattice
  data (solid lines). Individual components with average weighting
  $\sim 33<Q^2_i>^{-1/2}$ are shown in the upper and
  lower part by dot-dashed and dashed lines.} 
\label{fig:gluonprop}
\end{figure}

\end{document}